\documentstyle[12pt]{article}
\input{epsfig.sty}
\newcommand{\beq}{\begin{eqnarray}}
\newcommand{\eeq}{\end{eqnarray}}
\begin{document}
\title{D Production In p-p and d-Au Collisions}
\author{Leonard S. Kisslinger\\
Department of Physics, Carnegie Mellon University, Pittsburgh, PA 15213\\
Ming X. Liu and Patrick McGaughey\\
P-25, Physics Division, Los Alamos National Laboratory, Los Alamos, NM 87545}
\date{}
\maketitle
\begin{abstract}
This is an extension of our previous work on $J/\Psi$, $\Psi'(2S)$,
$\Upsilon(nS)$ production in p-p and A-A collisions to the production of
$D^+(c\bar{d}),D^o(c\bar{u})$, with the main new aspect being the
fragmentation probability, $D_{c \rightarrow c\bar{q}}$, which has been
calculated almost two decades ago. The rapidity cross sections for
$D^+(c\bar{d}),D^o(c\bar{u})$ production from both p-p and d-AU collisions
is estimated.
\end{abstract}
\noindent
PACS Indices:12.38.Aw,13.60.Le,14.40.Lb,14.40Nd
\vspace{1mm}

\section{Introduction}
  We consider $D^+(c\bar{d}),D^o(c\bar{u})$ production via unpolarized p-p 
collisions at 200 GeV, an extension of our previous work on $J/\Psi,\Psi'(2S)$,
and $\Upsilon(nS)$ production\cite{klm11}. In addition to being an important 
study of QCD, it also could provide a test of the production of Quark Gluon 
Plasma (QGP) in relativistic heavy ion collisions (RHIC), which would be an 
extension of our work on A-A production of heavy quark states\cite{klm14}. 
Estimates of $D^+(c\bar{d}),D^o(c\bar{u})$ production via d-Au collisions are 
also made, using the methods of Ref.\cite{klm14}. 

  As in our previous work we use the color octet model\cite{cl96,bc96,fl96},
which is consistent with experimental studies at E=200 GeV \cite{nlc03,cln04}. 
In Refs.\cite{klm11},\cite{klm14} the mixed hybrid theory was used for the
production of $\Psi'(2S),\Upsilon(3S)$, but this not relevant for the
present theory.

  The main new aspect of the present work is that while a gluon can produce a
$c\bar{c}$ or $b\bar{b}$ state, it cannot directly produce a $c\bar{d}$.
A fragmentation process converts a $c\bar{c}$ into a $c\bar{d}-d\bar{c}$,
for example. We use the fragmentation probability, 
$D_{c \rightarrow c\bar{q}}$ of Bratten et. al.\cite{bcfy95}.

\section{Differential $pp\rightarrow DX$ cross section}

Using what in Ref\cite{ns06} is called scenerio 2, the production
cross section with gluon dominance for DX is
\beq
\label{1}
  \sigma_{pp\rightarrow DX} &=&  \int_a^1 \frac{d x}{x} 
f_g(x,2m)f_g(a/x,2m) \sigma_{gg\rightarrow DX}\; ,
\eeq
with\cite{bcfy95}
\beq
\label{2}
 \sigma_{gg\rightarrow DX}&=& 2 \sigma_{gg\rightarrow c\bar{c}}
D_{c\rightarrow c\bar{q}} \; ,
\eeq
where $\sigma_{gg\rightarrow c\bar{c}}$ is similar to the charmonium production
cross section in Ref\cite{klm11} and $D_{c\rightarrow c\bar{q}}$ is the total
fragmentation probability. 
For $E=\sqrt{s}$=200 GeV the gluon distribution funtion is
\beq
\label{fg}
 f_g(y) &=& 1334.21 - 67056.5 x(y) + 887962.0 x(y)^2
\eeq

We use the quark fragmentation probability, 
$D_{c \rightarrow c\bar{q}}$ of Bratten et. al.\cite{bcfy95}, illustrated in the
figure below.

\begin{figure}[ht]
\begin{center}  
\epsfig{file=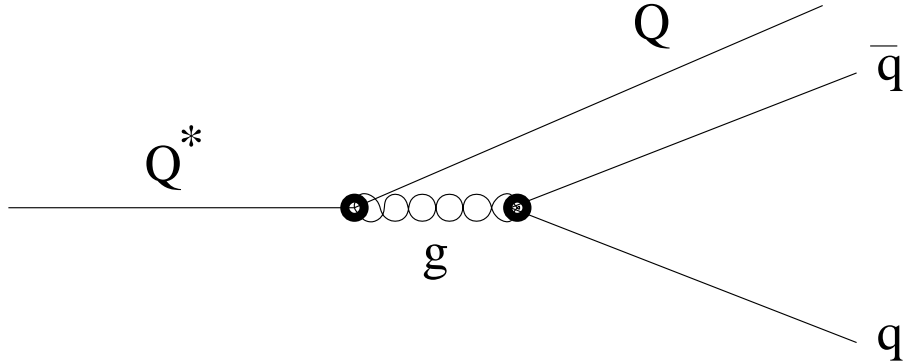,height=5 cm,width=12cm}
\caption{ Quark fragmentation for $Q^* \rightarrow (Q\bar{q})+q)$}
\label{}
\end{center}
\end{figure}

From Ref\cite{bcfy95}, using for the light quark mass=(up-mass+down-mass)/2=
3.5 Mev.
\beq
\label{3}
    D_{c \rightarrow c\bar{q}}&=& 9.21 \times 10^{5} \alpha_s^2 |R(0)|^2/\pi 
 \; ,
\eeq
in units of $(1/GeV^3)$, with $\alpha_s=.26$. For a 1S state
$|R(0)| ^2 = 4/(a_o)^3$. For a $c\bar{q}$ state, $(1/a_o)=m_q\simeq 3.5$ MeV. 
Therefore,
\beq
\label{R(0)}
   |R(0)| ^2&\simeq& 1.71 \times 10^{-7} {\rm \;\;(GeV)^3} \nonumber \\
    D_{c \rightarrow c\bar{q}}&\simeq& 3.39 \times 10^{-3} \; .
\eeq

The calculation of the cross section is similar to that in Ref\cite{klm11}.
\beq
\label{4}
  \frac{d\sigma_{pp\rightarrow DX}}{dy}&=& Acc*f_g(x(y),2m) f_g(a/x(y),2m) 
\frac{dx(y)}{dy} \frac{1}{x(y)} D_{c \rightarrow c\bar{q}} \; ,
\eeq
with rapidity $y$
\beq
\label{y(x)}
      y &=& \frac{1}{2} ln (\frac{E + p_z}{E-p_z}) \nonumber \\
        x(y) &=& 0.5 \left[\frac{m}{E}(\exp{y}-\exp{(-y)})+\sqrt{(\frac{m}{E}
(\exp{y}-\exp{(-y)}))^2 +4a}\right] \; , 
\eeq
where $Acc$ is the matrix element for charmonium production\cite{klm11} 
modified by an effective mass $ms$: $ Acc= 7.9*10^{-4} (1.5/ms)^3 nb$.

 From Eq(\ref{4}) we find $\frac{d\sigma_{pp\rightarrow DX}}{dy}$ shown in
the figure below, with $ms$=1.5 GeV.
\vspace{2cm}

\begin{figure}[ht]
\begin{center}
\epsfig{file=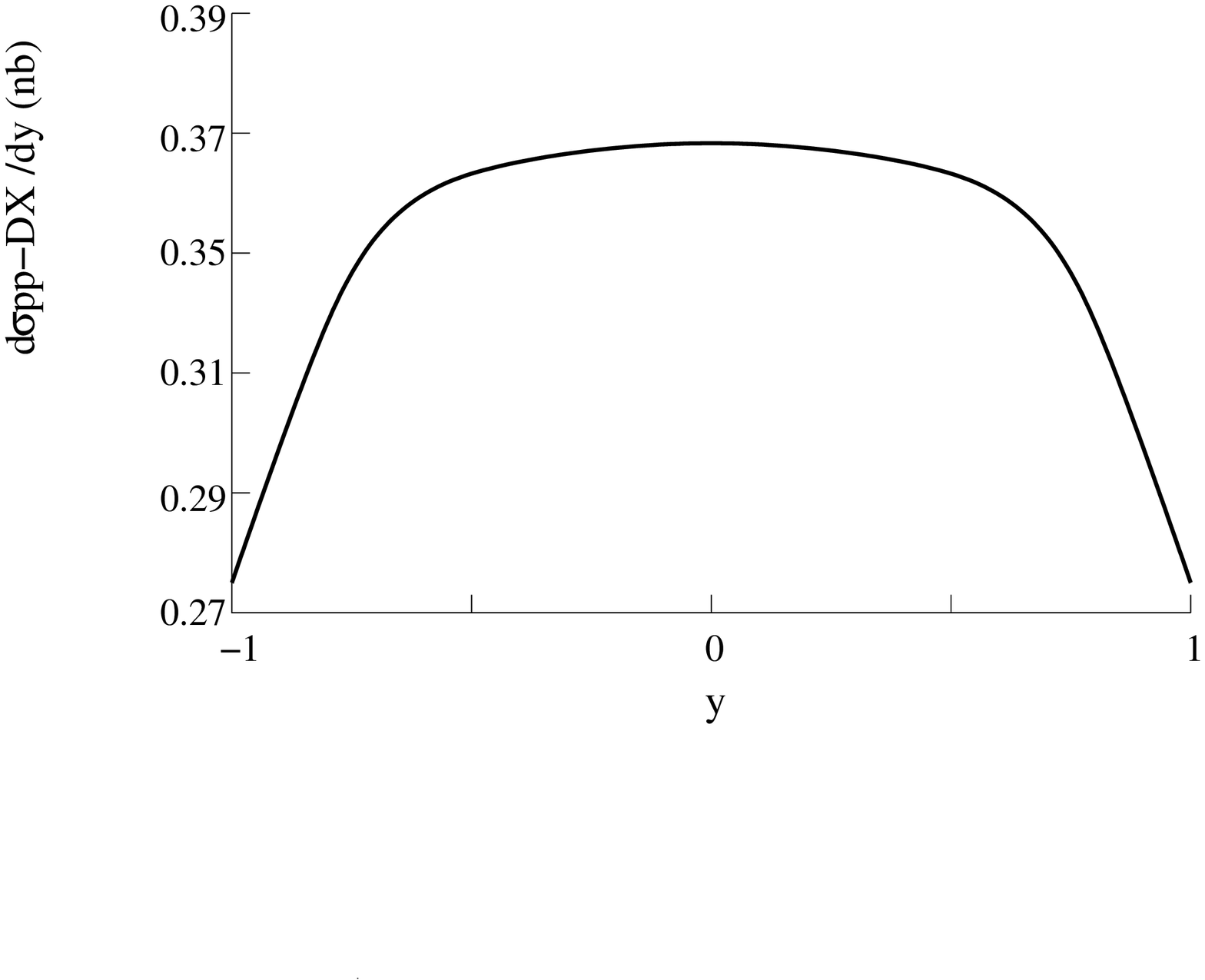,height=10cm,width=12cm}
\end{center}
\caption{$d\sigma/dy$ for E=200 GeV unpolarized p-p collisions producing D+X} 
\end{figure}

\section{Total $pp\rightarrow DX$ cross section}

The total cross section for $pp\rightarrow DX$ is\cite{klm11}
\beq
\label{5}
  \sigma_{pp\rightarrow DX}&=&  \int_a^1\frac{dx}{x} Acc*fg(x(y),2m) 
fg(a/x(y),2m)  D_{c \rightarrow c\bar{q}} \; .
\eeq

From Eqs(\ref{fg},\ref{R(0)}) and $Acc$ one obtains
\beq
\label{sigma}
  \sigma_{pp\rightarrow DX}&=& 2.678 {\rm \mu b}
\eeq
\newpage

  A number of experiments have measured $\sigma_{c\bar{c}}$ cross 
sections at $\sqrt{s_{pp}}$=200 GeV\cite{phenix06,phenix07,star07,phenix09}.
Theoretical estimates of heavy quark state production via p-p collisions
at RHIC and LHC energies were made almost two decades ago\cite{ggm95}.
Experimental measurements of $D^+, D^-, D^0$ production via p-p collisions 
are expected in the future.

\section{Differential $dAu\rightarrow DX$ cross section}

Open Charm yields in d+AU Colllisions at $\sqrt{s_{NN}}$=200 GeV have been 
measured via STAR\cite{star05} and PHENIX\cite{phenix14} experiments.
Cold nuclear matter effects on heavy-quark production were estimated
for a number of rapidities via PHENIX experiments\cite{phenix14}. We use 
the results of this experiment for the study of $D$ production via d-Au 
collisions.

In this Section we estimate the production of $D^+, D^0$ from d-Au collisions, 
using the methods given in Ref.\cite{klm14} for the estimate of production of 
$\Psi$ and $\Upsilon$ states via Cu-Cu and Au-Au collisions based on p-p 
collisions.

  The differential rapidity cross section for D+X production via d-Au
collisions is given by $ \frac{d\sigma_{pp\rightarrow DX}}{dy}$ with modification
described in Ref.\cite{klm14} for Cu-Cu and Au-Au collisions:
\beq
\label{sigmadAu}
  \frac{d\sigma_{dAu\rightarrow DX}}{dy}&=& R_{dAu}N^{dAu}_{coll}
\left (\frac{d\sigma_{pp\rightarrow DX}}{dy} \right)  \; ,
\eeq
where $R_{dAu}$ is the nuclear-modification factor, $N^{dAu}_{coll}$ is the
number of binary collisions, and $\left (\frac{d\sigma_{pp\rightarrow DX}}{dy} 
\right)$ is the differential rapidity cross section for $DX$ production via
nucleon-nucleon collisions in the nuclear medium.

  $\left (\frac{d\sigma_{pp\rightarrow DX}}{dy} \right)$ is given by Eq(\ref{4})
with $x(y)$ replaced by the function $\bar{x}$, the effective parton x in the 
nucleus Au\cite{vitev06}:
\beq
\label{barx}
         \bar{x}(y)&=& x(y)(1+\frac{\xi_g^2(A^{1/3}-1)}{Q^2}) \; ,
\eeq
which was evaluated in Ref.\cite{klm14}, where it was shown that 
$\bar{x}(y) \simeq x(y)$

  In Ref.\cite{phenix14} the  quantities  $R_{dAu}$ and $N^{dAu}_{coll}$
(called $R_{dA}$ and $<N_{coll}>$ in that article) were estimated from
experiments on p+p and d+Au collisions. From that reference (see FIG.3)
we use
\beq
\label{R-N}
        R_{dAu} &\simeq& 1.0 \nonumber \\
        N^{dAu}_{coll}&\simeq& 10.0 \; .
\eeq
Note that Ref.\cite{phenix14} shows $\simeq$ 50 \% larger $R_{dA}$ for
muons at negative rather than positive rapidity. However, the magnitude of
these rapidities is much larger than those in our calculation.

  From Eqs(\ref{4},\ref{sigmadAu},\ref{R-N}), one obtains the differential 
rapidity cross section for D+X production via dAu collisions,
 $ \frac{d\sigma_{dAu\rightarrow DX}}{dy}$ shown in Figure 3

\clearpage

\begin{figure}[ht]
\begin{center}
\epsfig{file=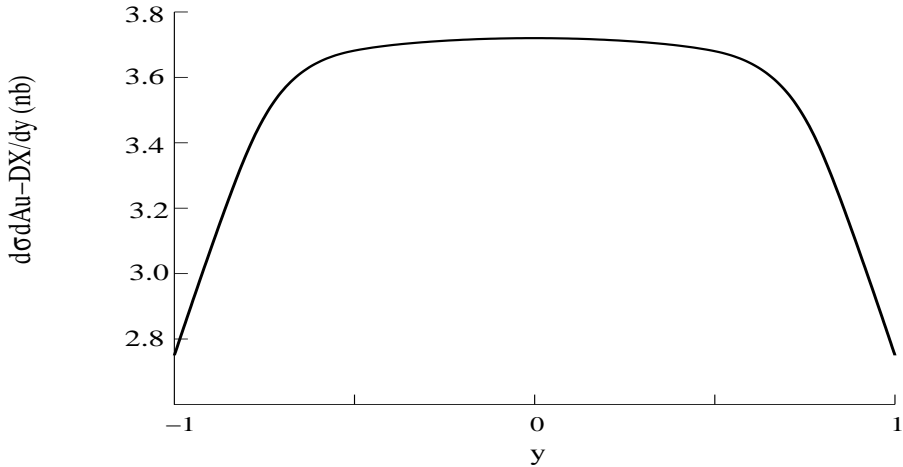,height=8cm,width=12cm}
\end{center}
\caption{$d\sigma/dy$ for E=200 GeV  d-Au collisions producing D+X} 
\end{figure}

  In Ref.\cite{star05} although $ \frac{d\sigma_{dAu\rightarrow DX}}{dy}$ was
not measured, the electron distributions for $p+p \rightarrow e+X$ and
$d+AU \rightarrow e+X$ were measured (Ref.\cite{star05}, FIG. 2) as a function 
of transverse momentum ($p_{T}$), and the ratio $d+AU \rightarrow e+X/p+p 
\rightarrow e+X$ in the range $1<p_{T}<2$ GeV/c was consistent with 
$R_{dAu}N^{dAu}_{coll}\simeq 10$, Eq(\ref{R-N}).

\section{Conclusions}

We have estimated the production of heavy-quark mesons 
$D^+(c\bar{d}),D^o(c\bar{u})$ +$X$ via p-p colllisions using the color
octet model with an extension of our previous work on production of
$\bar{c}c$ and $\bar{b}b$ states to $\bar{d}c$ or $\bar{u}c$ D-meson
states using fragmentation. Our results are expected to be tested by p-p 
collision experiments in the future. We have also estimated the production
of D-meson states via d-Au collisions, using experimental results
for the nuclear modification and number of binary collisions in recent
d-Au collisions experiments, which also might be measured in future
experiments. 

\vspace{1cm}
\Large{{\bf Acknowledgements}}\\
\normalsize
This work was supported in part by a grant from the Pittsburgh Foundation,
and in part by the DOE contracts W-7405-ENG-36 and DE-FG02-97ER41014.

\end{document}